# Anomalous thermal transport in metallic transition-metal nitrides originated from strong electron-phonon interactions

Shouhang Li[1#], Ao Wang[1#], Yue Hu[1], Xiaokun Gu[2], Zhen Tong[3] and Hua Bao[1*]

[1]University of Michigan-Shanghai Jiao Tong University Joint Institute, Shanghai Jiao Tong University, Shanghai 200240, China

[2]Institute of Engineering Thermophysics, School of Mechanical Engineering, Shanghai Jiao Tong University, Shanghai 200240, China

[3]Shenzhen JL Computational Science and Applied Research Institute, Shenzhen 518110, China

Metallic transition-metal nitrides (TMNs) are promising conductive ceramics for many applications, whose thermal transport is of great importance in device design. It is found metallic TiN and HfN hold anomalous thermal transport behaviors compared to common metals and nonmetallic TMNs. They have extremely large intrinsic phonon thermal conductivity mainly due to the large acoustic-optic phonon frequency gaps. The phonon thermal conductivity is reduced by two orders of magnitude as the phonon-isotope and phonon-electron scatterings are considered, which also induce the nontrivial temperature-independent behavior of phonon thermal conductivity. Nesting Fermi surfaces exist in both TiN and HfN, which cause the strong electron-phonon coupling strengths and heavily harm the transport of phonons and electrons. The phonon component takes an abnormally large ratio in total thermal conductivity, as 29% for TiN and 26% for HfN at 300 K. The results for thin films are also presented and it is shown that the phonon thermal conductivity can be efficiently limited by size. Our findings provide a deep understanding on the thermal transport in metallic TMNs and expand the scope of heat conduction theory in metal.

---

[*] To whom correspondence should be addressed. Email: hua.bao@sjtu.edu.cn (HB)

[#]S.L. and A.W. contribute equally to this work.

# I. Introduction

Metallic transition-metal nitrides (TMNs) exhibit superior properties of both metals and ceramics, such as extremely high melting points (>1800 K), wear resistance, high electrical conductivity, *etc.*[1]. Metallic TMNs have been widely used as in machine cutting tools [2], gate electrodes [3], thermal interfacial materials [4], *etc*. Especially, metallic TMNs are widely used to synthesize superlattices for high-temperature thermoelectric applications [5-8]. A high figure of merit (ZT) as 1.5 can be achieved by ZrN/ScN superlattice at 1300 K [5]. The extremely low thermal conductivity of metallic TMN-based superlattices has been believed to be the major reason for the excellent thermoelectric performance [5,6]. It is known that nonmetallic nitrides are usually good at heat conduction, for example, AlN holds high thermal conductivity of 321 W/mK at room temperature [9]. However, although electrons are involved in heat conduction, the metallic TMNs are known that the thermal conductivity is much lower, typically in the range of 3.8~63 W/mK [10,11]. Therefore, understanding the thermal transport mechanism in metallic TMNs is crucial to the further design the novel metallic TMN-based functional materials.

The understanding of thermal transport in metallic TMNs is still quite limited. Due to the difficulty in preparing defect-free samples, available thermal conductivities of metallic nitrides are scattered. Taking the widely used TiN as an example, the experimental thermal conductivity is located in a wide range of 19.2 [1] and 63 W/mK [11]. Other than obtaining the total thermal conductivity values, the separation of phonon and electron thermal conductivity is of great importance, especially for thermoelectrinc applications [5,6] and thermal interfacial materials [4,12]. It is tacitly believed that the phonon thermal conductivity contributes less than 10% [13], while recent experimental results suggest that the phonon thermal conductivity can still contribute ~35% of the thermal conductivity in VN at 300 K [14]. However, the experimental results are also questionable because they are generally obtained through subtracting the electronic component from total thermal conductivity [10,11,14,15] taking a certain Lorenz number, either assuming to be the Sommerfeld value, or estimated from the Bloch-Gruneisen model. This induces large uncertainty as the Lorenz number is known to deviate the Sommerfeld value [16] except for free-electron metal at high temperatures.

The investigation on the mechanism of heat conduction in metallic TMNs requires a deeper understanding in the scattering mechanism for both phonon and electron transport. The common understanding is the case that the phonon thermal conductivity in metals is mainly limited by anharmonic phonon-phonon interactions near the Debye temperature or higher, and the phonon-electron interactions only take effects at low temperatures [17-19]. Therefore, the phonon thermal conductivity ($\kappa_{ph}$) of metals is believed to follow the relationship of $\kappa_{ph} \sim 1/T$ [12] as the temperature is comparable to the Debye temperature. For electron thermal conductivity in metals, it is usually thought to be a constant at high temperatures [20,21]. As such, the total thermal conductivity of metals should be almost constant at high temperatures considering the small contribution of phonon. Unexpectedly, it is found the thermal conductivity of metallic TMNs shows a positive temperature coefficient [15,20]. As it is noted, phonon transport in metallic ceramics can be limited by both vacancies and conduction electrons [22,23]. However, the effect of conduction electrons on phonon transport in metallic TMNs is still unclear as it is easy to be concealed by the effect of defects [24]. Meanwhile, the electron transport is strongly related to the scatterings from phonons and defects. As thus, the separate analysis on the electron phonon interactions is a key step to understand the anomalous behavior of thermal conductivity with temperature in metallic TMNs.

In this work, we focus on the thermal transport mechanism in the metallic TMNs. Two typical metallic nitrides TiN and HfN are studied by a rigorous first-principles analysis on both phonon and electron transport. We first carry out the calculation on the bulks, and then it is expanded to thin film of the same materials. The findings are expected to be helpful in interpreting the exotic phenomena in experiments and improve the performance of metallic TMN-based functional materials.

## II. Methods and Simulation Details

Including the contributions of both phonons and electrons, the total thermal conductivity of metallic TMNs is written as $\kappa_{tot} = \kappa_{ph} + \kappa_{el}$. In this work, both the phonon and electron transport are studied at the mode level based on the framework of Boltzmann transport equations (BTE). The phonon thermal conductivity tensor is expressed as [25,26]

$$\kappa_{ph,\alpha\beta} = \frac{1}{N_{\mathbf{q}}} \sum_{\lambda} c_{v,\lambda} v_{\lambda,\alpha} v_{\lambda,\beta} \tau_{\lambda} \qquad (1)$$

where $N_{\mathbf{q}}$ is the number of $\mathbf{q}$-points in the First Brillouin zone. $\lambda = (\mathbf{q}, \nu)$ is the phonon mode denotation. $c_{v,\lambda}$ is the phonon heat capacity, $v_{\lambda}$ is the phonon group velocity, and $\tau_{\lambda}$ is the phonon relaxation time. According to Matthiessen's rule, the phonon scattering rate can be expressed as $1/\tau_{\lambda} = 1/\tau_{\lambda}^{ph\text{-}ph} + 1/\tau_{\lambda}^{ph\text{-}iso} + 1/\tau_{\lambda}^{ph\text{-}el}$. Here, $1/\tau_{\lambda}^{ph\text{-}ph}$, $1/\tau_{\lambda}^{ph\text{-}iso}$ and $1/\tau_{\lambda}^{ph\text{-}el}$ are phonon scattering rates induced by three-phonon, phonon-isotope and phonon-electron interactions. The detailed description for these three phonon scattering rates can be found in the Supplemental Material S1. Other phonon scatterings, like phonon-boundary and phonon-vacancy scatterings are not included in this work.

Electron is the other heat carrier in metallic nitrides. For electrical transport properties, we adopt the calculation scheme proposed in our previous work [16], i.e. momentum relaxation time for electrical conductivity and energy relaxation time for thermal conductivity. The electrical conductivity ($\sigma^{\alpha\beta}$) and thermal conductivity ($\kappa_{el}^{\alpha\beta}$) tensor can be expressed as

$$\begin{cases} \sigma^{\alpha\beta} = -\dfrac{q^2}{N_{\mathbf{k}}V}\sum_{m\mathbf{k}} v_{m\mathbf{k}}^{\alpha} v_{m\mathbf{k}}^{\beta} \tau_{m\mathbf{k}}^{\sigma}(\mu,T) \dfrac{\partial f^0(\varepsilon_{m\mathbf{k}},\mu,T)}{\partial \varepsilon_{m\mathbf{k}}} \\ \kappa_{el}^{\alpha\beta} = \dfrac{1}{N_{\mathbf{k}}V}\sum_{m\mathbf{k}} -\dfrac{(\varepsilon_{m\mathbf{k}}-\mu)^2}{T} v_{m\mathbf{k}}^{\alpha} v_{m\mathbf{k}}^{\beta} \tau_{m\mathbf{k}}^{\kappa}(\mu,T) \dfrac{\partial f^0(\varepsilon_{m\mathbf{k}},\mu,T)}{\partial \varepsilon_{m\mathbf{k}}} \end{cases} \quad (2)$$

$q$ is the elementary charge. $N_{\mathbf{k}}$ is the total number of **k**-points in the first Brillouin zone. $\alpha$ and $\beta$ are Cartesian coordinate components. $v_{m\mathbf{k}}$ is the electron group velocity. $\tau^{\sigma}$ and $\tau^{\kappa}$ are the momentum and energy relaxation time for electron, respectively. The expressions for the two electron relaxation times can be found in Ref [16].

Metallic TMNs-based thin films are widely used in experiments and we calculate both electron and phonon cross-plane thermal conductivity for thin films by solving the BTE [27,28]. We set the thermalizing boundaries with $T_1$ and $T_2$ on the two surfaces of the thin film. The cross-plane thermal conductivity is defined as $\kappa = qL/(T_1-T_2)$ [28]. $L$ is the thickness of the thin film, and $Q$ is the heat flux, which is expressed as:

$$Q = \sum_{\lambda} \frac{1}{3} c_{v,\lambda} v_{\lambda} \Lambda_{\lambda} S(\Lambda_{\lambda},L)(T_1-T_2)/L. \quad (3)$$

The mean free path $\Lambda_{\lambda}$ of each electron/phonon can be obtained from first-principles calculations described above. $S(\Lambda_{\lambda},L)$ is the suppression function, and the details can be founded in Ref. [27]. This model is accurate for the isotropic material, which is the case for metallic nitrides TiN and HfN.

All the first-principles calculations are performed with Quantum Espresso [29]. The Perdew-Burke-Ernzerhof (PBE) form [24] of the exchange-correlation functional is employed. The cutoff energy of the plane wave is set as 80 Ry, and the convergence threshold of electron energy is set to be $10^{-10}$ Ry for the self-consistent field calculation. The optimized lattice constants come out to be 4.258 Å and 4.512 Å (experimental values are 4.24 Å and 4.52 Å [1]) for TiN and HfN with face-centered cubic rock-salt lattice, respectively. The harmonic force constant is obtained employing the density-functional perturbation theory under $6\times6\times6$ **q**-points mesh. The cubic force constant is extracted with THIRDORDER.PY package [26]. A supercell of $4\times4\times4$ was used and the 4$^{th}$ nearest neighbors are included. A $3\times3\times3$ **k**-points mesh is used. The three-phonon scattering rates are obtained by solving the Peierls-BTE [25,26] using an iterative approach, while the other scatterings are included only within the relaxation time approximation. For the electronic properties calculations, the electron scattering rates are calculated by Electron-Phonon Wannier (EPW) package [30]. The electron-phonon coupling matrix elements are first calculated on the coarse grids of $6\times6\times6$ **k**-points and $6\times6\times6$ **q**-points, and are then interpolated to the dense mesh of $80\times80\times80$ **k**-points and $40\times40\times40$ **q**-points to ensure the convergence of electrical transport coefficients in the whole temperature range of 200-1000 K.

## III. Results and Discussions

### A. Phonon thermal conductivity

#### A.1. Anomalously large intrinsic phonon thermal conductivity

First, we examine the phonon thermal conductivity limited by intrinsic three-phonon scatterings. Both TiN and HfN hold ultrahigh intrinsic phonon thermal conductivity, shown by black lines in Fig. 1. Specifically, the phonon thermal conductivity of HfN is 3335 W/mK at 300 K, which is higher than reported BAs (3170 W/mK) and close to diamond (3450 W/mK) [31]. HfN holds smaller average phonon group velocity compared to TiN, while it holds much larger average phonon relaxation time (shown in Supplemental Material S4), which results in its higher phonon thermal conductivity compared to TiN. In contrast to the large phonon thermal conductivity of metallic TMNs presented here, the rock-salt nonmetallic TMNs are found to hold low phonon thermal conductivity smaller than 100 W/mK at 300 K [32]. The ultrahigh thermal conductivity of metallic TMNs is related to the unique features in their phonon dispersions shown in Fig. 2 (a) and (b). The intrinsic three-phonon scattering includes the following processes, three acoustic phonons (*aaa*), two acoustic phonons and an optic phonon (*aao*), and an acoustic phonon and two optic phonons (*aoo*). Firstly, the extremely large phonon frequency gap strongly impedes the *aao* processes. Then, bunching behavior of acoustic phonon braches near the phonon peculiarities (a dip and softening of acoustic modes, shown by orange arrows in Fig. 2) decrease the phase space for *aaa* scattering processes [33]. Narrow optic phonon bandwidth furtherly impedes the *aoo* processes [34]. These three factors make the three-phonon scattering rates abnormally small, shown in Fig. 3. In contrast to metallic TMNs here, nonmetallic TMNs hold small phonon thermal conductivity as their phonon dispersions lack the three characteristics mentioned above [32].

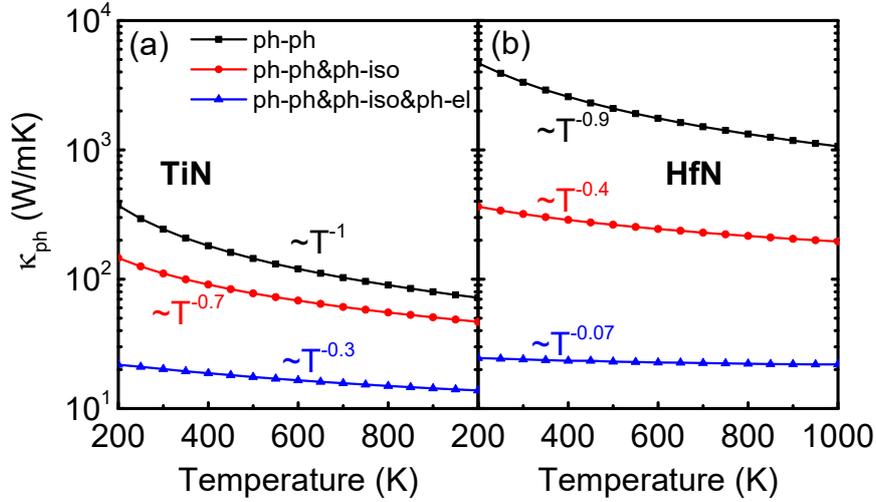

FIG. 1. The calculated phonon thermal conductivity limited by combination of phonon-phonon (ph-ph), phonon-isotope (ph-iso), and phonon-electron (ph-el) interactions as a function of temperature for (a) TiN and (b) HfN.

Large acoustic-optic (*ao*) phonon frequency gaps exist in the phonon dispersions in Fig 2 (a) and (b), which is related to the large atomic mass difference between metallic atoms and the nitrogen atom. The acoustic phonon modes are mainly related to metallic atoms vibrations and N atoms vibrations for optical phonon modes, shown in the phonon density of states (DOS) in Fig. S1. HfN holds larger phonon frequency gap compared to TiN due to the heavier atomic mass of Hf. As we replace the atomic mass of Ti for the atomic mass of Hf in TiN, the frequency gap of TiN approaches that of HfN. A much higher phonon thermal conductivity of 1650 W/mK at 300 K is obtained compared to original TiN (243 W/mK). Interestingly, phonon peculiarities appear in both TiN and HfN from high symmetry paths Γ to X and K to Γ, marked by the orange arrows in Fig. 2 (a) and (b). They are signs for nesting Fermi surfaces [35], which are shown in Fig. 2 (c) and (d) and will be discussed later.

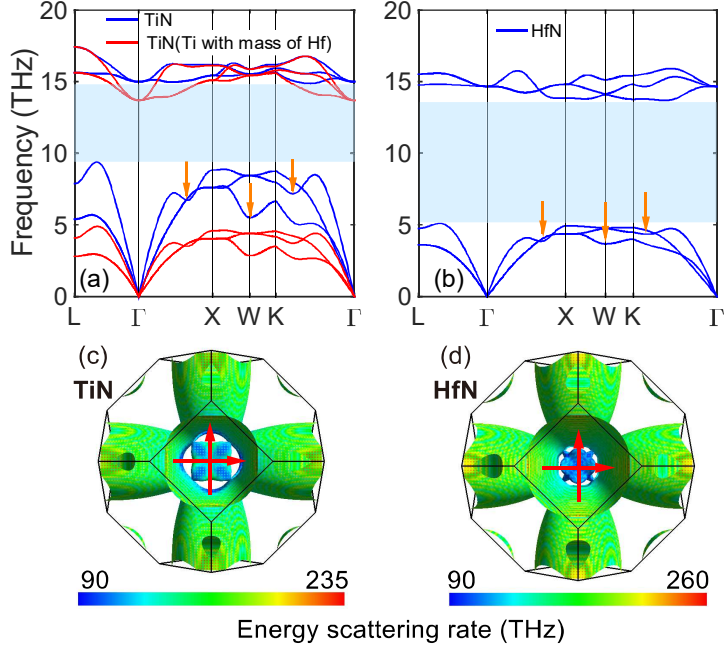

FIG. 2. Phonon dispersion curves along high-symmetry paths for (a) TiN and (b) HfN. The orange arrows show the phonon peculiarities for TiN and HfN. The large phonon frequency gaps are denoted by shallow blue ribbons. A larger phonon frequency gap is obtained in TiN as the atomic mass of Ti being replaced by atomic mass of Hf (red lines). Fermi surfaces showing the energy relaxation time ($\tau_{m\mathbf{k}}^{\kappa}$) at electron mode $n\mathbf{k}$ at 300 K for (c) TiN and (d) HfN. The red arrows indicate the nesting vectors. Fermi surfaces are plotted with FermiSurfer package [36].

**A.2. Pronounced effects from phonon-isotope scatterings**

The isotopic atoms are inevitable in the natural situation and would introduce phonon-isotope scatterings. Interestingly, the isotopic purification of atoms in metallic nitrides can significantly harm the phonon transport in metallic nitrides, shown by the red lines in Fig. 1. This is contradictory to the observations in nonmetallic nitrides, whose isotope effects are negligible [32]. For HfN, the phonon thermal conductivity changes one order of magnitude (from 3335 W/mK to 319 W/mK) as the isotope scattering is included. Here the factor $P = 100 \times \left( \kappa_{\text{ph,pure}} / \kappa_{\text{ph,natural}} - 1 \right)$ is employed to evaluate the isotope effects [37]. We obtain the value as 120 (945) for TiN (HfN) at room temperature. This is significantly larger than the previous report for other restrictive solids [37]. The isotope effects stem from the isotope atoms mass disorder [38], which can be quantified by the mass variance factors $g_i$ in Eq. (S2). The factors are $g_{Ti} = 2.86 \times 10^{-4}$, $g_{Hf} = 5.25 \times 10^{-5}$ and $g_{N} = 1.87 \times 10^{-5}$ for natural Ti, Hf, and N atoms, respectively. As shown by the shallow blue scatters in Fig. 3, the isotope scatterings are smaller than three-phonon scatterings in most frequency regions for TiN, while they are comparable in the range of 4-7 THz. The phonons in this frequency range have a large contribution of 51% to the intrinsic phonon thermal conductivity, shown in Fig. S2. For HfN, the phonon-isotope scatterings are evidently larger than three-phonon scattering rates in the entire phonon frequency range. Since the isotope scattering is static and temperature-independent, the phonon thermal conductivities become less temperature-dependent, shown by the red lines in Fig 1.

**A.3. Extremely strong phonon-electron scatterings**

In metallic TMNs, abundant conductive electrons exist in the crystals [39,40], which is the essential difference from other electrical resistive ceramics. Naturally, phonons can be scattered by conductive electrons and block the phonon transport. The quantitative calculations show that phonon-electron scattering is still much smaller than phonon-phonon scatterings in common metals [41-43]. Unexpectedly, the phonon thermal conductivities are significantly decreased as the phonon-electron scatterings are considered in TiN and HfN, shown by the blue lines in Fig. 1. Similar to the factor $P$, we adopt the factor $E = 100 \times (\kappa_{ph,int} / \kappa_{ph,int+ext} - 1)$ to quantify the extrinsic phonon scatterings (including phonon-isotope and phonon-electron scattering) effects on phonon thermal conductivity, with $\kappa_{ph,int}$ considering only intrinsic phonon scatterings and $\kappa_{ph,int+ext}$ considering both intrinsic and extrinsic phonon scatterings. The values would be 1106 (13812) for TiN (HfN), which suggest the extremely strong extrinsic phonon scattering effects. Since the phonon-electron scattering is also weakly temperature-dependent (shown in Supplemental Material S5), the final phonon thermal conductivity of TiN and HfN is almost temperature-independent, which goes to be $\kappa_{ph} \sim T^{-0.3}$ for TiN and $\kappa_{ph} \sim T^{-0.07}$ for HfN, respectively. This is distinct from nonmetallic TMNs [32], whose phonon thermal conductivity is strongly temperature-dependent as it is only limited by intrinsic phonon-phonon scatterings. This is also totally different from the observations in common conductive metals [41,42], where the phonon-electron scattering is relatively weak compared to phonon-phonon scattering. The phonon-electron scattering rates at 300 K are shown by the blue scatters in Fig. 3. For TiN, the phonon-electron scattering rates are the largest of all the scattering sources in the frequency range of 4~7 THz. For HfN, the phonon-electron scattering rates are larger than those of other scattering sources in almost the whole frequency range. Although the acoustic phonon-electron scattering rates of HfN is lower than these of TiN, the phonon thermal conductivity of HfN is reduced by phonon-electron scatterings much heavier due to the relatively small three-phonon scattering rates.

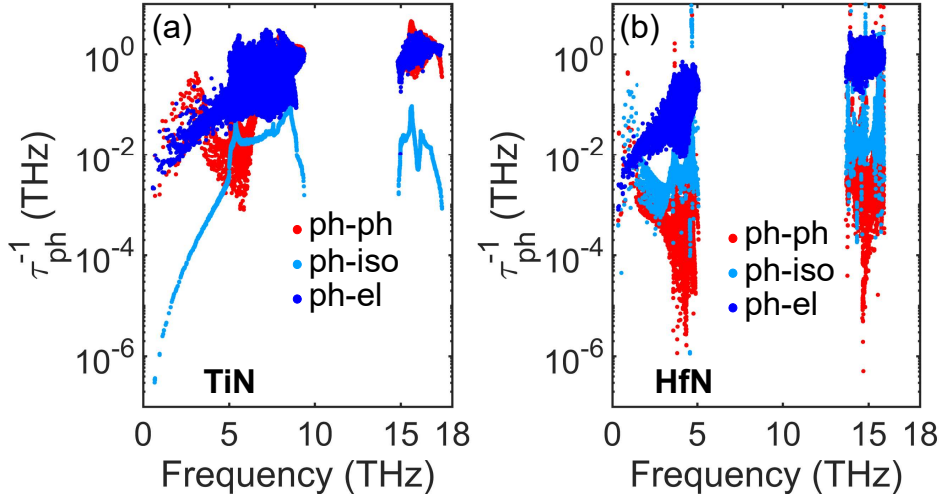

FIG. 3. The phonon scattering rates induced by phonon-phonon (ph-ph), phonon-isotope (ph-iso) and phonon-electron (ph-el) scattering at 300 K for (a) TiN and (b) HfN.

To further investigate the phonon-electron scattering mechanisms in TiN and HfN, phonon-electron scattering rates superimposed on acoustic phonon modes are shown in Fig. 4. The optic phonon modes are ignored as they almost have no contribution to the thermal conductivity, shown in Fig. S2. It can be observed that the phonon modes around the phonon peculiarities hold significantly larger phonon-electron scattering rates. The impeding effects on the phonon heat conduction through phonon-electron interactions are very efficient as these phonon modes near the peculiarities are the main contributor to phonon thermal conductivity. The phonon-electron scattering mechanism in the two metallic TMNs is totally different from that in other common metals, which do not have phonon peculiarities [41-43]. It can be speculated that strong phonon-electron scatterings may also exist in other metallic nitrides, like ZrN, whose phonon dispersion also hold peculiarities [35].

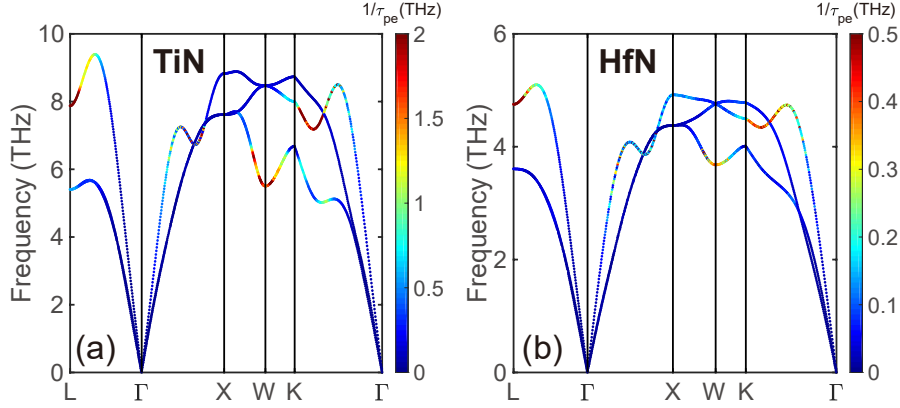

FIG. 4. Room-temperature phonon-electron scattering rates superimposed on the acoustic phonon dispersion curves along high symmetry paths for (a) TiN and (b) HfN.

**B. Electrical transport coefficients**

**B.1. Relatively small electron thermal conductivity**

The electrons are the other important heat carriers in metallic TMNs. They are mainly scattered by phonons at elevated temperatures. The electronic thermal conductivity is shown in Fig. 5. It is almost temperature invariant in the temperature range of 200-1000 K. The electronic heat capacity is almost linearly increased with temperature, while electron relaxation time is linearly decreased with temperature at elevated temperatures. These two competitive factors lead to the temperature-independent behavior of the electronic thermal conductivity. TiN (HfN) holds relatively small electronic thermal conductivity, as 49 W/mK (69 W/mK) at 300 K compared to common metals [41,43-45]. The small electronic thermal conductivity makes the phonon component non-negligible, shown by the blue ribbons in Fig. 5. The phonon component takes a ratio of 29% (26%) in total thermal conductivity for TiN (HfN) at 300 K. This is much larger than the phonon contributions in other common metals [41,43]. Our calculated total thermal conductivity for TiN agrees well with the measured data for high-quality stoichiometric epitaxial TiN reported in Ref. [11], as 63 W/mK at 300 K. The value presented here is significantly larger than the data reported in Ref. [1], as 19.2 W/mK. It should stem from defects in experimental samples which strongly impede the heat conduction.

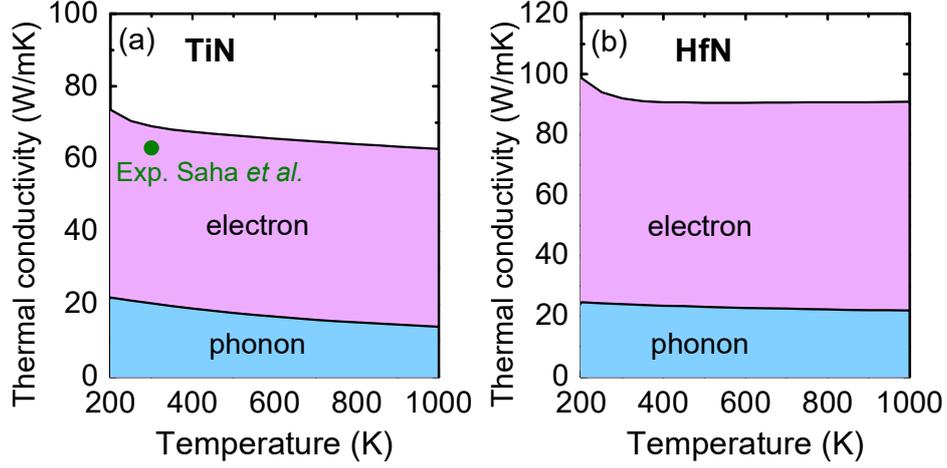

FIG. 5. The calculated electron and phonon thermal conductivity as a function of temperature for (a) TiN and (b) HfN. The blue and purple ribbons are the phonon and electron contribution to the total thermal conductivity, respectively. The experimental data of TiN is taken from Ref. [11].

**B.2. Electrical conductivity and Wiedemann-Franz law**

The electrical conductivity decreases in inverse proportion with the temperature mainly due to the decrease of electron relaxation time, shown in Fig. 6 (a) and (c). It can be seen that the calculated results match the experimental data well. The electrical conductivity of both TiN and HfN is on the magnitude of $10^6$ S/m, which is significantly larger than the dielectric ceramics, while lower than the common elementary metals [46]. This is attributed to the relatively small electron relaxation time, which is strongly limited by phonons. With thermal conductivity and electrical conductivity obtained, the Lorenz ratio can be evaluated by $L = \kappa / \sigma T$, shown in Fig. 6 (b) and (d). As we only consider the electronic thermal conductivity, the Lorenz number would first increase and then tend to converge to almost a constant value at higher temperatures, where the large angle scattering dominants in electron scattering [16]. The Lorenz number estimated by the total thermal conductivity would be overestimated in the whole temperature range, due to the non-negligible phonon contribution to the total thermal conductivity. Wiedemann-Franz law is often employed to evaluate the electron thermal conductivity [10,14]. Either Sommerfeld value $L_0 = 2.44 \times 10^{-8} \, W\Omega/K^2$ or Lorenz number estimated by Bloch- Grüneisien model (See Supplemental Material S6) is used and they have departure from the value estimated by mode-level calculation, shown in Fig. 6 (b) and (d). The electron thermal conductivity estimated by mode-level calculation, Wiedemann-Franz law using Sommerfeld value and Bloch-Grüneisen model are listed in Table S2. There is significant difference in electron thermal conductivity estimated by mode-level calculation and Wiedemann-Franz law. The uncertainties in electron thermal conductivity estimated through Wiedemann-Franz law will be transported to phonon thermal conductivity, which is indirectly obtained by subtracting the electronic component from total thermal conductivity.

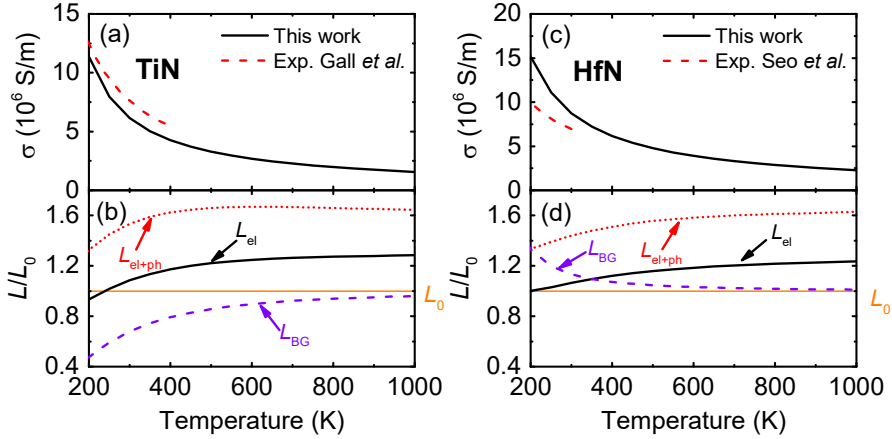

FIG. 6. The calculated phonon-limited electrical conductivity as a function of temperature for (a) TiN and (c) HfN. The experimental data for TiN and HfN are taken from Ref. [40] and Ref. [39], respectively. The calculated Lorenz number as a function of temperature for (b) TiN and (d) HfN. The Lorenz numbers estimated by Bloch-Grüneisen model ($L_{BG}$) and Lorenz numbers calculated with ($L_{el+ph}$) and without ($L_{el}$) phonon thermal conductivity are normalized by Sommerfeld value, as $L_0 = 2.44 \times 10^{-8} \, W\Omega/K^2$. The orange solid lines mark the Sommerfeld value $L_0$.

The small electrical and thermal conductivity are attributed to the extremely large electron-phonon coupling constants, as 0.65 (TiN) and 0.74 (HfN). The anomaly strong electron-phonon coupling stems from the unique Fermi surfaces in metallic TMNs. Nesting Fermi surfaces exist in both TiN and HfN, shown in Fig. 2 (c) and (d). There are six arms with large parallel regions and providing abundant scattering sites for specific phonon modes. The electron energy scattering rates are superimposed on the Fermi surface and it is clear to see the electron modes at the end of the arms hold larger scattering rates, implying strong electron-phonon coupling in these regions. Simultaneously, phonons would be strongly scattered by the electrons at these sites and results in anomalously large phonon-electron scatterings. In contrast to previous work [47], the strong Fermi surface nesting exists in the Group IVb transition-metal form nitrides, rather than the Group Vb transition-metal form carbides.

### C. Thermal transport in thin films

Due to the outstanding stability of metallic TMNs, they are usually employed as the base materials in thin films or superlattices [6,48]. Here, the thermal conductivity in the cross-plane of thin films with thickness of 50, 100, and 500 nm is shown in Fig. 7. Both phonon and electron thermal conductivity increase with thickness of thin films. The electron thermal conductivity is bulk-like as the thickness approaches 500 nm. The size effect on electron thermal conductivity is very weak as the temperature is larger than 600 K, while it is significant on phonon thermal conductivity in the whole temperature range. The phonon thermal conductivity of bulk TiN (HfN) is even reduced by ~42% (~27%) with a large thickness of 500 nm at 300 K. It is attributed to the different thermal conductivity accumulation functions of phonon/electron thermal conductivity with respect to phonon/electron mean free path (MFP), shown in Fig. S5. The phonon MFP is about one order of magnitude larger than that of electron, which is opposite to the observation in common elementary metals [41]. As thus, the phonon thermal conductivity is more efficiently limited by size compared to electron thermal conductivity in metallic TMNs.

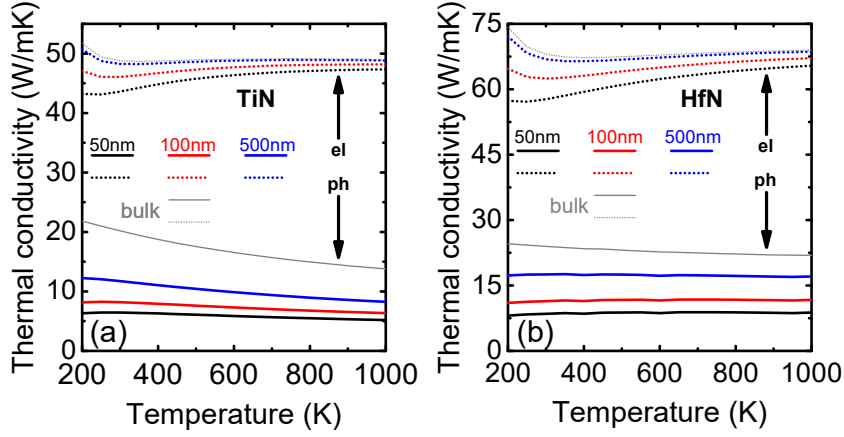

FIG. 7 The cross-plane phonon thermal conductivity (solid lines) and electron thermal conductivity (dashed lines) with a thickness of 50 (black lines), 100 (red lines) and 500 nm (blue lines) for (a) TiN and (b) HfN. The bulk values for phonon (grey solid lines) and electron (grey dash lines) are also presented.

## IV. Conclusions

In summary, the thermal transport mechanisms in metallic TMNs are investigated through first-principles calculations. Two typical metallic nitrides TiN and HfN are studied as prototypes. They hold totally different heat conduction behaviors compared to common conductive metals and nonmetallic TMNs. It is found they have extremely large intrinsic phonon thermal conductivity mainly attributed to the large acoustic-optic phonon frequency gaps. The anomalously large phonon-isotope and phonon-electron scattering are observed which can reduce the intrinsic phonon thermal conductivity by two orders of magnitude. The unexpected strong electron-phonon coupling strength is related to the nesting Fermi surface which heavily limits both phonon and electron transport. The phonon thermal conductivity is nearly temperature-independent due to the temperature invariant phonon-isotope and phonon-electron scatterings. The phonon contribution to total thermal conductivity is non-negligible owing to the relatively small electronic component. The size effects on electron thermal conductivity are weak in thin films at high temperatures, while the phonon thermal conductivity can be efficiently limited by size in the whole temperature range.

**Acknowledgments**

We would like to thank Dr. Cheng Shao for valuable discussion. H.B. acknowledges the support by the National Natural Science Foundation of China No. 51676121 (HB). X. G. acknowledges the support by the National Natural Science Foundation of China No. 51706134. Simulations were performed with computing resources granted by HPC (π) from Shanghai Jiao Tong University.

**References**

[1] H. O. Pierson, *Handbook of Refractory Carbides & Nitrides: Properties, Characteristics, Processing and Apps* (William Andrew, 1996).

[2] P. Hedenqvist, M. Olsson, P. Wallén, Å. Kassman, S. Hogmark, and S. Jacobson, Surface and Coatings Technology **41**, 243 (1990).

[3] H. Yu *et al.*, IEEE electron device letters **24**, 230 (2003).

[4] R. M. Costescu, M. A. Wall, and D. G. Cahill, Physical Review B **67**, 054302 (2003).

[5] M. Zebarjadi, Z. Bian, R. Singh, A. Shakouri, R. Wortman, V. Rawat, and T. Sands, Journal of electronic materials **38**, 960 (2009).

[6] P. Eklund, S. Kerdsongpanya, and B. Alling, Journal of Materials Chemistry C **4**, 3905 (2016).

[7] J. L. Schroeder, D. A. Ewoldt, R. Amatya, R. J. Ram, A. Shakouri, and T. D. Sands, Journal of microelectromechanical systems **23**, 672 (2013).

[8] B. Saha, A. Shakouri, and T. D. Sands, Applied Physics Reviews **5**, 021101 (2018).

[9] Z. Cheng *et al.*, Physical Review Materials **4**, 044602 (2020).

[10] W. S. Williams, Jom **50**, 62 (1998).

[11] B. Saha, Y. R. Koh, J. P. Feser, S. Sadasivam, T. S. Fisher, A. Shakouri, and T. D. Sands, Journal of Applied Physics **121**, 015109 (2017).

[12] A. Majumdar and P. Reddy, Applied Physics Letters **84**, 4768 (2004).

[13] T. M. Tritt, *Thermal conductivity: theory, properties, and applications* (Springer Science & Business Media, 2005).

[14]	Q. Zheng, A. B. Mei, M. Tuteja, D. G. Sangiovanni, L. Hultman, I. Petrov, J. E. Greene, and D. G. Cahill, Physical Review Materials **1**, 065002 (2017).

[15]	R. E. Taylor and J. Morreale, Journal of the American Ceramic Society **47**, 69 (1964).

[16]	S. Li, Z. Tong, X. Zhang, and H. Bao, arXiv preprint arXiv:2004.08843 (2020).

[17]	A. Sommerfeld and H. Bethe, Verlag Julius Springer, Berlin **1** (1933).

[18]	R. Makinson, in *Mathematical Proceedings of the Cambridge Philosophical Society* (Cambridge University Press, 1938), pp. 474.

[19]	P. Klemens, Australian Journal of Physics **7**, 57 (1954).

[20]	W. S. Williams, Journal of the American Ceramic Society **49**, 156 (1966).

[21]	Y. S. Touloukian, Thermophysical properties of matter **1** (1970).

[22]	L. Radosevich and W. S. Williams, Physical Review **181**, 1110 (1969).

[23]	D. T. Morelli, Physical Review B **44**, 5453 (1991).

[24]	J. P. Perdew, K. Burke, and M. Ernzerhof, Physical review letters **77**, 3865 (1996).

[25]	D. A. Broido, M. Malorny, G. Birner, N. Mingo, and D. A. Stewart, Applied Physics Letters **91**, 231922 (2007).

[26]	W. Li, J. Carrete, N. A. Katcho, and N. Mingo, Computer Physics Communications **185**, 1747 (2014).

[27]	C. Hua and A. J. Minnich, Journal of Applied Physics **117**, 175306 (2015).

[28]	Y. Hu, T. Feng, X. Gu, Z. Fan, X. Wang, M. Lundstrom, S. S. Shrestha, and H. Bao, Physical Review B **101**, 155308 (2020).

[29]	P. Giannozzi *et al.*, Journal of physics: Condensed matter **21**, 395502 (2009).

[30]	S. Poncé, E. R. Margine, C. Verdi, and F. Giustino, Computer Physics Communications **209**, 116 (2016).

[31]	L. Lindsay, D. A. Broido, and T. L. Reinecke, Physical review letters **111**, 025901 (2013).

[32]     C. Li and D. Broido, Physical Review B **95**, 205203 (2017).

[33]     L. Lindsay, D. Broido, and T. Reinecke, Physical review letters **111**, 025901 (2013).

[34]     S. Mukhopadhyay, L. Lindsay, and D. S. Parker, Physical Review B **93**, 224301 (2016).

[35]     E. I. Isaev, S. I. Simak, I. Abrikosov, R. Ahuja, Y. K. Vekilov, M. Katsnelson, A. Lichtenstein, and B. Johansson, Journal of applied physics **101**, 123519 (2007).

[36]     M. Kawamura, Computer Physics Communications **239**, 197 (2019).

[37]     L. Lindsay, D. A. Broido, and T. L. Reinecke, Physical review letters **109**, 095901 (2012).

[38]     S.-i. Tamura, Physical Review B **30**, 849 (1984).

[39]     H. S. Seo, T. Y. Lee, J. G. Wen, I. Petrov, J. E. Greene, and D. Gall, Journal of Applied Physics **96**, 878 (2004).

[40]     D. Gall, I. Petrov, and J. Greene, Journal of Applied Physics **89**, 401 (2001).

[41]     A. Jain and A. J. McGaughey, Physical Review B **93**, 081206 (2016).

[42]     Y. Wang, Z. Lu, and X. Ruan, Journal of applied Physics **119**, 225109 (2016).

[43]     Z. Tong, S. Li, X. Ruan, and H. Bao, Physical Review B **100**, 144306 (2019).

[44]     A. Giri, J. T. Gaskins, L. Li, Y.-S. Wang, O. V. Prezhdo, and P. E. Hopkins, Physical Review B **99**, 165139 (2019).

[45]     A. Giri, M. Tokina, O. Prezhdo, and P. Hopkins, Materials Today Physics, 100175 (2020).

[46]     J. Bass, Electrical Resistivity, Kondo and Spin Fluctuation Systems, Spin Glasses and Thermopower", eds. K.-H. Hellwege and JL Olsen, Landolt-Börnstein Numerical Data and Functional Relationships in Science and Technology, New Series, Group III: Crystal and Solid State Physics **15**, 1 (1982).

[47]     C. Li, N. K. Ravichandran, L. Lindsay, and D. Broido, Physical review letters **121**, 175901 (2018).

[48] P. E. Hovsepian, D. Lewis, and W.-D. Münz, Surface and Coatings Technology **133**, 166 (2000).

# Anomalous thermal transport in metallic transition-metal nitrides originated from strong electron-phonon interactions

Shouhang Li[1#], Ao Wang[1#], Yue Hu[1], Xiaokun Gu[2], Zhen Tong[3] and Hua Bao[1*]

[1]University of Michigan-Shanghai Jiao Tong University Joint Institute, Shanghai Jiao Tong University, Shanghai 200240, China

[2]Institute of Engineering Thermophysics, School of Mechanical Engineering, Shanghai Jiao Tong University, Shanghai 200240, China

[3]Shenzhen JL Computational Science and Applied Research Institute, Shenzhen 518110, China

## S1. The phonon scattering rates from different scattering sources

The intrinsic phonon transport is impeded by the three-phonon [1,2] scattering rate, which can be obtained following Fermi's "golden rule" as

$$\frac{1}{\tau_\lambda^{\text{ph-ph}}} = 2\pi \sum_{\lambda_1 \lambda_2} \left| V_{\lambda \lambda_1 \lambda_2} \right|^2 \left[ \begin{array}{l} \frac{1}{2}\left(1 + n_{\lambda_1}^0 + n_{\lambda_2}^0\right) \delta\left(\omega_\lambda - \omega_{\lambda_1} - \omega_{\lambda_2}\right) \\ + \left(n_{\lambda_1}^0 - n_{\lambda_2}^0\right) \delta\left(\omega_\lambda + \omega_{\lambda_1} - \omega_{\lambda_2}\right) \end{array} \right] \quad \text{(S1)}$$

where $V_{\lambda \lambda_1 \lambda_2}$ is three-phonon and four-phonon matrix element, respectively. $\delta$ is the Dirac delta function which ensures the conservation of energy during the scattering process.

---

* To whom correspondence should be addressed. Email: hua.bao@sjtu.edu.cn (HB)
#S.L. and A.W. contribute equally to this work.

There are also extrinsic phonon scattering sources, including phonon-isotope interactions and phonon-electron interactions. The phonon-isotope scattering is induced by the natural isotopic atoms and can be estimated by Tamura theory [3]

$$\frac{1}{\tau_\lambda^{\text{ph-iso}}} = \frac{\pi}{2}\omega_\lambda^2 \sum_{i \in \text{u.c.}} g_i \left|\mathbf{e}_{\lambda'}^*(i) \cdot \mathbf{e}_\lambda(i)\right|^2 \delta(\omega_\lambda - \omega_{\lambda'}) \quad \text{(S2)}$$

where $\mathbf{e}_\lambda$ is the normalized eigenvector of phonon mode $\lambda$ and the asterisk denotes the complex conjugate. The mass variance factor is expressed as $g_i = \sum_j f_i(j)\left[1 - m_i(j)/\bar{m}_i\right]^2$, with $f_i(j), m_i(j)$, and $\bar{m}_i$ the concentration, atomic mass of the $j$th substitution atom and average mass of the $i$th atom.

The phonon-electron scattering is induced by the conductive electrons which can be calculated from the imaginary part of phonon self-energy $\Pi_\lambda$ by $1/\tau_\lambda^{pe} = 2\,\text{Im}(\Pi_\lambda)/\hbar$. It can be explicitly expressed as [4]

$$\frac{1}{\tau_\lambda^{\text{ph-el}}} = \frac{2\pi}{\hbar} \sum_{mn} \int_{\text{BZ}} \frac{d\mathbf{k}}{\Omega_{\text{BZ}}} \left|g_{mn}^\nu(\mathbf{k},\mathbf{q})\right|^2 \left[f_{n\mathbf{k}}^0(\mu,T) - f_{m\mathbf{k}+\mathbf{q}}^0(\mu,T)\right] \delta(\varepsilon_{m\mathbf{k}+\mathbf{q}} - \varepsilon_{n\mathbf{k}} - \omega_\lambda) \quad \text{(S3)}$$

where $\Omega_{\text{BZ}}$ is the volume of the first Brillouin zone. $m$ and $n$ are the band index of electrons. $g$ is the electron-phonon matrix element. $f^0$ is Fermi-Dirac distribution. $\mu$ is the Fermi energy. $\varepsilon$ is the energy of the electron. The electron-phonon matrix element quantifies the electron with the initial state of $n\mathbf{k}$ approaches to the final state $m\mathbf{k}+\mathbf{q}$ and it is expressed as

$$g_{mn}^\nu(\mathbf{k},\mathbf{q}) = \sqrt{\frac{\hbar}{2\omega_\lambda}} \langle \psi_{m\mathbf{k}+\mathbf{q}} | \nabla_\lambda V | \psi_{n\mathbf{k}} \rangle \quad \text{(S4)}$$

with $\psi$ the ground-state Bloch wave function and $|\nabla_\lambda V|$ the first-order derivative of the Kohn-Sham potential with respect to the phonon displacement.

## S2. The total and projected phonon density of state (DOS)

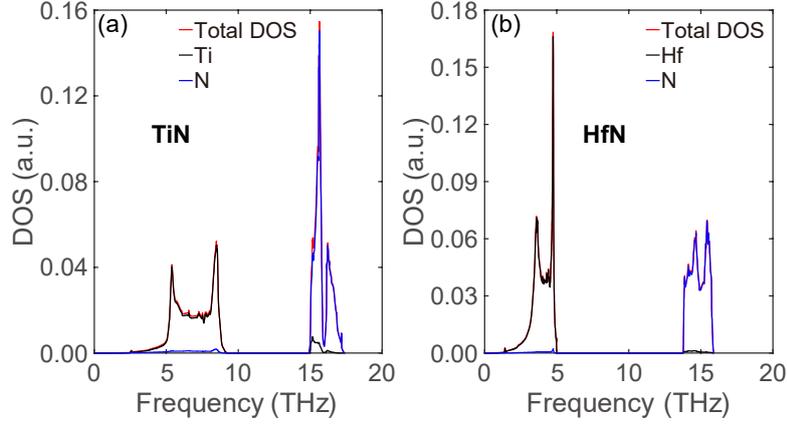

FIG. S1. The total and projected phonon DOS for (a) TiN and (b) HfN.

## S3. Spectral dependence of phonon thermal conductivity

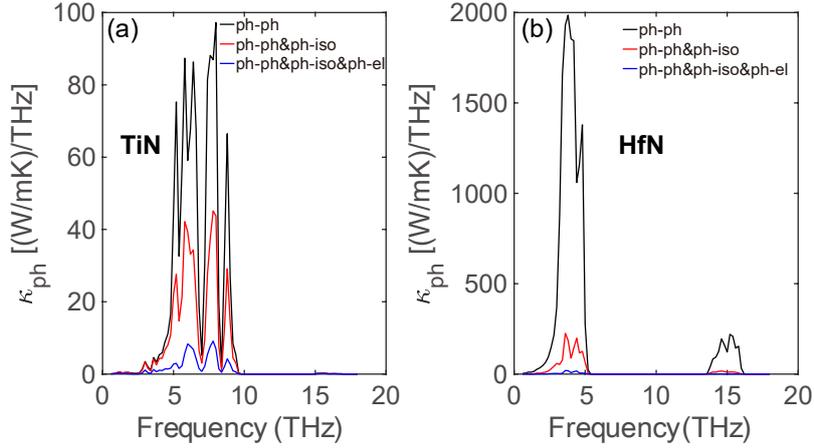

FIG. S2. Room-temperature frequency spectral dependence of phonon thermal conductivity limited by combination of phonon-phonon (ph-ph), phonon-isotope (ph-iso), and phonon-electron (ph-el) interactions at 300 K for (a) TiN and (b) HfN.

## S4. The average phonon relaxation time

The average phonon velocity and relaxation time are defined as following [5],

$$\begin{cases} \overline{v}_\alpha = \sqrt{\dfrac{\sum_\lambda c_{v,\lambda} v_{\lambda,\alpha} v_{\lambda,\alpha} \tau_\lambda}{\sum_\lambda c_{v,\lambda} \tau_\lambda}} \\ \overline{\tau}_\alpha = \dfrac{\sum_\lambda c_{v,\lambda} v_{\lambda,\alpha} v_{\lambda,\beta} \tau_\lambda}{\sum_\lambda c_{v,\lambda} v_{\lambda,\alpha} v_{\lambda,\alpha}} \end{cases} \quad (S5)$$

The average phonon group velocity is 1992 (576) m/s for TiN (HfN). The average three-phonon relaxation time is 52.23 (2027.5) ps for TiN (HfN) at 300 K.

**S5. The phonon-electron scattering rates with different temperatures**

The Eq. (S3) can be further simplified to be the following "double Delta approximation" form

$$\frac{1}{\tau_\lambda^{pe}} = \frac{2\pi}{\hbar} \omega_\lambda \sum_{mn} \int_{BZ} \frac{d\mathbf{k}}{\Omega_{BZ}} \left| g_{mn}^v(\mathbf{k},\mathbf{q}) \right|^2 \delta(\varepsilon_{n\mathbf{k}} - \varepsilon_F) \delta(\varepsilon_{m\mathbf{k}+\mathbf{q}} - \varepsilon_F), \quad (S6)$$

which is temperature-independent. The phonon-electron scattering rates at 300 K and 1000 K are shown in Fig. S3.

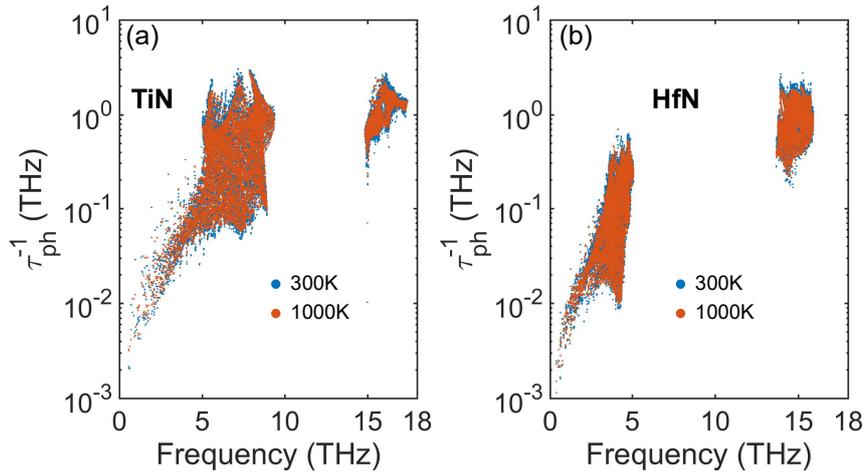

FIG. S3. Phonon-electron scattering rates at 300 K and 1000 K for (a) TiN and (b) HfN.

**S6. Lorenz number estimated by the Bloch-Grüneisen model**

The Bloch- Grüneisen model gives the Lorenz number as [6]

$$L = \frac{L_0}{1 + \frac{3}{\pi^2}\left(\frac{k_F}{q_D}\right)^2\left(\frac{\theta_D}{T}\right)^2 - \frac{1}{2\pi^2}\frac{J_7(\theta/T)}{J_5(\theta/T)}} \tag{S7}$$

where $k_F$ and $q_D$ are Fermi wave vector and Debye wave vector, respectively. $k_F$ is calculated from electron concentration $n$, as $k_F = (3\pi^2 n)^{1/3}$ and we adopt the concentration value reported in Ref. [7] for TiN and Ref. [8] for HfN. $q_D$ is calculated as $q_D = \frac{k_B \theta_D}{\hbar v_s}$, with $v_s$ the sound velocity. $\theta_D$ is Debye temperature which can be estimated from phonon DOS. $J_n$ is defined as

$$J_n\left(\frac{\theta}{T}\right) \equiv \int_0^{\theta/T} \frac{x^n e^x}{(e^x - 1)^2} dx \tag{S8}$$

with $n$ an integer. The parameters used in Bloch-Grüneisen model are listed in Table S1.

Table S1. The parameters used in Bloch-Grüneisen model

| Materials | Fermi wave vector (1/Å) | Debye wave vector (1/Å) | Debye temperature (K) |
|---|---|---|---|
| TiN | 1.1085 | 1.9764 | 817.29 |
| HfN | 0.5219 | 2.8316 | 690.86 |

The calculated Lorenz number is $0.6795 L_0$ for TiN and $1.1337 L_0$ for HfN, with Sommerfeld value $L_0 = 2.44 \times 10^{-8} \text{W}\Omega/\text{K}^2$.

The electron thermal conductivity calculated by mode-level method, Wiedemann-Franz law using Sommerfeld value and Bloch-Grüneisen model at 300 K is shown in Table S2. The electrical conductivity used in Wiedemann-Franz law is taken from Ref. [7] for TiN and Ref. [8] for HfN.

Table S2. The electron thermal conductivity calculated by different methods

| Materials | Mode-level calculation (W/mK) | Wiedemann-Franz law | |
|---|---|---|---|
| | | Sommerfeld value (W/mK) | Bloch-Grüneisen model (W/mK) |
| TiN | 48.79 | 56.36 | 38.30 |
| HfN | 68.02 | 50.87 | 57.68 |

## S7. The electron-phonon coupling constant

The electron-phonon coupling constant is denoted as [9]

$$\lambda = 2\int_0^\infty \frac{\alpha^2 F(\omega)}{\omega} d\omega \quad (S9)$$

where $\alpha^2 F(\omega)$ is the Eliashberg spectral function, which is expressed as

$$\alpha^2 F(\omega) = \frac{1}{N(\varepsilon_F)} \sum_\lambda \sum_{\mathbf{k}mn} |g_{mn}^\nu(\mathbf{k},\mathbf{q})|^2 \delta(\hbar\omega - \hbar\omega_\lambda) \times \delta(\varepsilon_{\mathbf{k}n} - \mu) \delta(\varepsilon_{\mathbf{k+q}m} - \mu) \quad (S10)$$

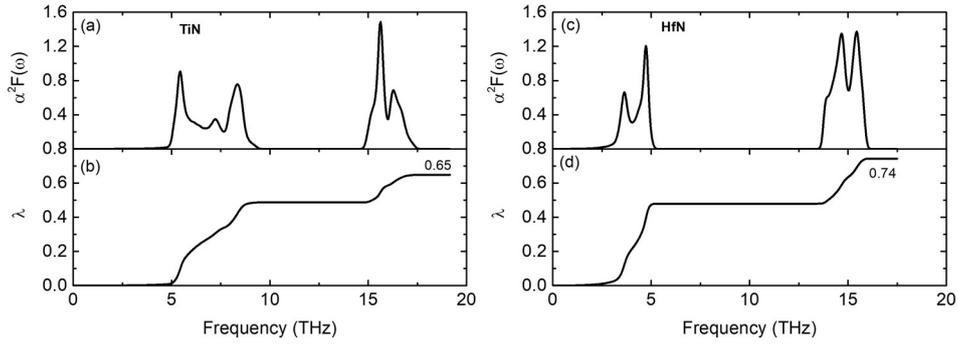

FIG. S4. The transport spectral function for (a) TiN and (c) for HfN. The electron-phonon coupling constant for (b) TiN and (d) HfN. The acoustic phonon modes contribute most to the electron-phonon scattering.

## S8. Thermal conductivity accumulation functions with respect to phonon/electron MFP

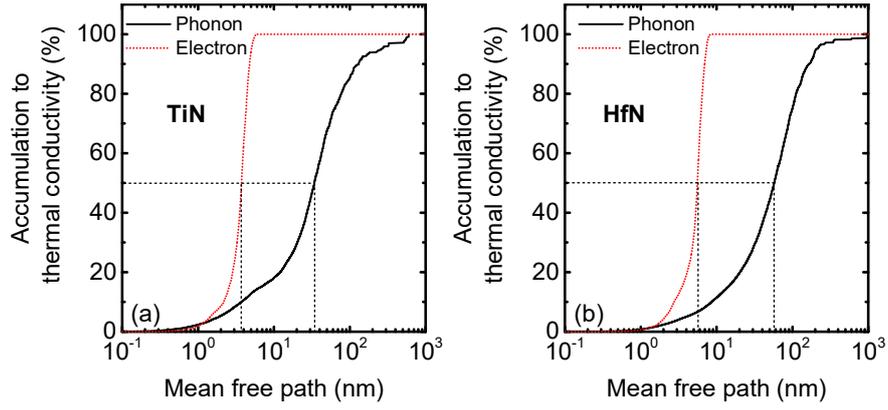

FIG. S5. The accumulated phonon/electron thermal conductivity with respect to phonon/electron mean free path (MFP) at 300 K for (a) TiN and (b) HfN. The phonon MFP is about one order of magnitude larger than that of electron.

[1]  M. Omini and A. Sparavigna, Physical Review B **53**, 9064 (1996).
[2]  D. A. Broido, M. Malorny, G. Birner, N. Mingo, and D. A. Stewart, Applied Physics Letters **91**, 231922 (2007).
[3]  S.-i. Tamura, Physical Review B **27**, 858 (1983).
[4]  S. Poncé, E. R. Margine, C. Verdi, and F. Giustino, Computer Physics Communications **209**, 116 (2016).
[5]  R. Guo, X. Wang, Y. Kuang, and B. Huang, Physical Review B **92**, 115202 (2015).
[6]  T. M. Tritt, *Thermal conductivity: theory, properties, and applications* (Springer Science & Business Media, 2005).
[7]  D. Gall, I. Petrov, and J. Greene, Journal of Applied Physics **89**, 401 (2001).
[8]  H. S. Seo, T. Y. Lee, J. G. Wen, I. Petrov, J. E. Greene, and D. Gall, Journal of Applied Physics **96**, 878 (2004).
[9]  F. Giustino, Reviews of Modern Physics **89**, 015003 (2017).